# PREDICTING THE 2020 US PRESIDENTIAL ELECTION WITH TWITTER

Michael Caballero

Divison of Computing, Data Science, and Society,
University of California, Berkeley, Berkeley, CA, USA

## ABSTRACT

*One major sub-domain in the subject of polling public opinion with social media data is electoral prediction. Electoral prediction utilizing social media data potentially would significantly affect campaign strategies, complementing traditional polling methods and providing cheaper polling in real-time. First, this paper explores past successful methods from research for analysis and prediction of the 2020 US Presidential Election using Twitter data. Then, this research proposes a new method for electoral prediction which combines sentiment, from NLP on the text of tweets, and structural data with aggregate polling, a time series analysis, and a special focus on Twitter users critical to the election. Though this method performed worse than its baseline of polling predictions, it is inconclusive whether this is an accurate method for predicting elections due to scarcity of data. More research and more data are needed to accurately measure this method's overall effectiveness.*

## 1. INTRODUCTION

Over the course of this past year, I have been conducting independent research to predict the winner of the 2020 presidential election. The specific topic is estimating the public opinion of candidates throughout the 2020 presidential election using social media data. Due to the difficulty of collecting social media data, this project has been simplified to only use Twitter data. The main question for this project is: can social media replicate or outperform the accuracy of national polls in predicting the vote share of the 2020 US presidential election? This these is idea stems from an interest that I have always had in politics and government. This research allows me to combine my data science skill set with my passion and informal knowledge of politics.

The use of data in political campaigns has been publicly highlighted in recent years, and research on effectively utilizing data is quite relevant to the future of political science. A robust and repeatable methodology for polling support of campaigning politicians would significantly affect campaign strategies; a method with similar accuracy to traditional polling techniques could potentially replace those systems. In some races, like those for state senators or governors, it could be the only reliable method for polling as traditionally these races do not have accurate polling data. Nevertheless, prediction with social media runs into other significant obstacles already tackled with traditional polling platforms and techniques. For instance, traditional polling specifically accounts for geographical and demographical issues. In particular, Gayo-Avello mentions how traditional polling methods accounts for noise like spam tweets, self-selection bias, likely voters, and demographical biases [1]. Classical polling focuses heavily on these methods to create a strategy for measuring public opinion that emphasizes only including opinions from prospective voters, removing information not related to the election, and correcting any bias based on the demographic bias of those surveyed. While the field of polling through social media has high aspirations, there are still many important challenges to overcome.

More accurate polling insights would help parties decide which races are highly competitive and thus where precious party funds could be spent most effectively. Pouring party resources and funding into a tight race is an effective strategy for political parties. And, this method could even be important for races with extensive polling data. Traditional polling methods are often expensive and time-intensive so a cheap, efficient model that relieson social media data would provide valuable complementary data. In sum, Twitter and other social media platforms can provide a huge sample size of data that could allow cheaper polling in real-time.

While the specific case of electoral prediction using social media data is interesting and would be an important contribution to political science, this research is also important for all social sciences. This research shows the potential for social media prediction to impact political science by allowing campaigns, analysts, and politicians to better understand voters and their political support. Specifically, it gives a glimpse into potential future techniques that could facilitate this understanding. But it also furthers social science by contributing to a new method of prediction based on social media data. This type of prediction foreshadows a transition in social sciences as these disciplines are undergoing a shift from data scarcity to abundance. As more interactions happen on digital platforms, the capacity of social scientists to predict the attitudes and behaviors of society should grow. Thus, this field of electoral forecasting using social media data is a representation of a larger transition in the social sciences due to growing data availability.

## 2. ETHICAL ISSUES

As of now, I do not see any prominent ethical issues with applying this research to political campaigns. There is a small possibility that developing a better method of polling could have negative impacts on groups without a fair share of political power. Better methods of polling would likely first be utilized by politicians in power. These politicians have the most connections and resources and often are the first to adopt new successful campaign techniques. It could give the powerful an unfair advantage in influencing democratic outcomes thus having negative repercussions on those marginalized or excluded from the political sphere like people with disabilities or religious and ethnic minorities (National Democratic Institute). I find this ethical concern to be small because this is only one possibility of developing better polling methods. This development could also be beneficial to those without power by evening the playing field of politics -- making accurate polling data cheaper and thus more accessible to all. As it is difficult to predict how any new technology will affect society, these ethical concerns should be considered though should not prevent progress in this particular research.

One bigger ethical question with this broader subject of research, using social media to measure public opinion, is the use of social media data without the consent of its users. Yes, social media users do sign Terms of Service before utilizing these platforms but it is safe to assume most users do not understand how their data is being used. Third parties' use of social media data is quite extensive and further research and thought are needed to determine whether this type of research is actually ethical.

## 3. LITERATURE REVIEW

Gayo-Avello's meta-analysis on electoral prediction from Twitter data is the first meta-analysis published on the subject and one of the best introductory resources to gain a better understanding of the literature [1]. While this meta-analysis is slightly outdated, it still relates to the current state of the field and provides a great snapshot of the accomplishments and challenges of early research. It mainly analyzes ten different studies that attempted to predict 16 different elections. This paper suggests that any approach to predict elections from Twitter data should be judged through four aspects: data collected, approach to deal with noise, methods of prediction, and overall evaluation. "Noise" in this context is mainly referring to tweets containing disinformation, propaganda, and rumors, or those originating from spammers or robots that obscure the regular interactions on the platform.

While all four aspects discussed are important, the two that are critical to understanding the field are methods of prediction and evaluation. The data collected has varied with each project and noise has mainly been ignored in recent research likely because it amplifies sentiments already on the platform. At this point, the research only focused on two methods of prediction for inferring votes: counting tweets and sentiment analysis. Tweet counting is a simple method that counts the number of mentions for a candidate or party assuming that the vote is strongly correlated with this number. Sentiment analysis focuses on scoring tweets as positive or negative, then aggregating these scores to determine electoral outcomes. On evaluation, he points out that there is little consensus in the literature currently and thus there should be consistent ways to report and compare findings. Gayo-Avello summarizes three main ways to evaluate electoral predictions. First, predicting the winner; did the method predict the correct outcome? Second, what is the mean absolute error (MAE) between predicted vote share and actual vote share? Lastly, he suggests measuring the correlation with pre-electoral polls using metrics like $R^2$ or root mean squared error. Importantly, Gayo-Avello suggests using an election baseline to provide more context to the accuracy of reported results. He asserts that MAE "does not allow for comparison neither across papers nor across races" and that underperforming a baseline "should be considered unsuccessful" [1].

Yet Gayo-Avello does heavily criticize the field of research as a whole, arguing that: "(1) baselines chosen to evaluate performance up to now are not realistic; (2) simple methods achieve inconsistent results when replicated; and (3) their presumed tolerance to noise should not be taken for granted but much better substantiated" [1]. And up to this point, common sense would agree as the field of electoral prediction with Twitter data had researchers often concluding different results and disagreeing. His last remarks accurately note that for a method to gain credibility it must correspond to a series of elections and should be generalized for different elections in different countries.

Skoric et al. provide an updated perspective of electoral and public opinion forecasting using social media data in the only other meta-analysis assessing the predictive power of social media data [2]. Skoric et al. first discuss the importance of the field of measuring public opinion; they counter critics of this field saying that "dismissing social media data as being invalid due to its inability to represent a population misses capturing the dynamics of opinion formation" [2]. In their analysis, the starting point for comparing methods of measuring public opinion is in examining the diversity of methods. The two main methods in the literature are categorized as either sentiment-based or structural-based. Sentiment methods have only slightly evolved since their definition in the previous meta-analysis -- utilizing lexicon-based or machine learning approaches to capture positive or negative sentiments and predict vote share. Conversely, previously defined tweet counting methods have evolved into this broad category of structural analyses. These methods utilize metrics of social networks like mentions, likes, and replies, to encompass the relational connections of the social media platform. Most often in sentiment analysis, researchers will generally sum the number of tweets multiplied by their sentiment and use the proportion of each candidate to represent vote share. In structural analysis, proportion of vote share is simply represented by proportion of total mentions or likes. Combining additional sources of data beyond one social media source is complex with different techniques depending on the literature. Skoric et al. also analyze these different social media sources that literature in this field utilizes, providing insights into what platforms could be the best for measuring public opinion. Lastly, they compare the differences in methodology and data sources using a combination of MAE and $R^2$. Similar to GayoAvello[1], Skoric et al. emphasize the need for "a more standardized way of reporting data collection methods and statistical estimates of predictive power" [2].

The results of this meta-analysis provide future researchers hints at what methods would work best in measuring the public's political opinion. This meta-analysis focused on 437 estimates from 74 different research studies. In these comparisons between data and methods, they emphasize that $R^2$ be the primary metric for comparison as $R^2$ and similar correlation metrics "manifested low variance and thus are more stable across different studies" [2]. They found that structural approaches, with a mean $R^2$ of 0.605, perform substantially better than sentiment approaches, mean $R^2$ of 0.492. Interestingly, they also measured studies that combined structural and sentiment approaches and found these to have the highest average $R^2$, 0.621. For social media data sources, there were a number of different platforms compared including Twitter, Facebook, forums, blogs, and Youtube. Of these platforms, Twitter ranked second-best by MAE, but fourth-best by $R^2$; blogs performed best by both of these metrics.

Tumasjan et al. is perhaps the most influential piece of literature in this field with many scholars agreeing that this article started this line of research [3]. It used the 2009 federal election in Germany to "investigate whether Twitter is used as a forum for political deliberation and whether online messages on Twitter validly mirror offline sentiment" [3]. In practice, this study used over 100,000 tweets that mentioned a party or politician in the election and performed two separate analyses, one volumetric and one sentiment-based. They concluded that the sentiment analysis reflected some of the nuances of the campaign and found that just the number of messages reflects the election result and was close to replicating the accuracy of traditional election polls.
Jungherr et al. [4] and Metaxas et al. [5] provided a quick criticism to Tumasjan et al..The first study by Jungherr et al. repeated the analysis performed by Tumasjan et al. and found that their claims were unwarranted for a number of reasons. Mainly, Tumasjan et al. failed to specify well-grounded rules for data collection, the choice of parties to track, and the correct period. Jungherr et al. found a larger MAE and concluded that Tumasjan et al. made an incorrect prediction when taking into account all parties running for the election. Conversely, Metaxas et al. tried to duplicate the findings of Tumasjan et al. by performing a volume and sentiment analysis on several senate races. This study concluded that these methods for analyzing Twitter data are slightly better than chance for predicting elections. And Metaxas et al. suggested the use of a baseline, finding that those methods "were not competent compared to the trivial method of predicting through incumbency" [5].

Bermingham and Smeaton used the Irish general election to investigate how to mine political sentiment through social media data [6]. Collecting tweets that mainly mentioned a party name or abbreviation, this study utilized both volume and sentiment-based analyses to measure vote share. They concluded that Twitter does display a predictive quality with the volume analysis being a stronger indicator than sentiment analysis. But,Gayo-Avello does find that this method, while predictive, "was not competitive against traditional polls" [1].

Jürgens et al. investigated political communication on Twitter by extracting a directed network of user interactions [7]. The study found that it exhibits small-world properties with the most well-connected nodes having the strongest influence on information dissemination on the platform. Interestingly, Jürgens et al. concluded that Twitter political communication is "highly dependent on a small number of users, critically positioned in the structure of the network" [7].

Shi et al. analyzed millions of tweets over the course of months leading up to the 2011 republican primary elections [8]. This study analyzed tweets collected over a six-month period primarily using a volumetric analysis with and without a sentiment analysis. The researchers also included a geographic analysis measuring the public opinion in certain states training their linear regression model on the RealClearPolitics (RCP) aggregate polling numbers for certain states. While Shi et al. did accurately predict trends for half of the candidates, they did not accurately replicate patterns in the polling of the other half of the field.

MacWilliams studied 2012 US senate races utilizing fan participation and mobilization metrics from Facebook to make electoral predictions [9]. The two metrics this paper focused on was likes on a candidates Facebook page and Facebook's "people talking about this" (PTAT) metric which counts interactions between users and a candidate's page.This paper concluded that their Facebook model was a better predictor than aggregated polls of outcomes in the senate races for five of the eight weeks they predicted the vote share. Though unrelated, this paper is one of the first to indirectly implement Jürgens et al.'s [7] findings that political communication in social networks is dependent on critical users. This shows that measuring the reaction or popularity of a candidate on social media could be a key factor in accurate electoral prediction.

Cameron et al. examined the ability of social media to predict election results in the 2011 New Zealand general election using both Facebook and Twitter [10]. This study mainly tracked the size of the 453 candidates' social media networks by collecting the number of friends a candidate has on Facebook and the number of followers a candidate had on Twitter. They found that "the social media are often statistically significant, but the size of their effects is small" thus concluding that "social media only has a practical effect in elections that are likely to be closely fought" [10].

Mirowski et al. investigated whether tweets created during the 2016 US presidential campaign could make accurate predictions about changes in a candidate's poll score trends [11]. This study collected 12 million tweets that mentioned candidates and tweets directly from candidates' accounts over the course of 60 days. For prediction, Mirowski et al. utilized a multivariate time series analysis to figure out whether temporal modeling has greater predictive power than attribute-based modeling. As this paper researched into whether the polling trend could be accurately classified as increasing or decreasing, it is difficult to compare their evaluation metrics with most of the other literature in the field. But, they did conclude that time-series models outperformed traditional attribute-based models.

Vepsäläinen et al. attempted to measure public opinion in the 2015 Finnish parliamentary elections using Facebook likes [12]. They collected data on 2.7 million Facebook likes directly from 2146 candidates' Facebook pages to estimate vote share for the 200 seats in the Finnish parliament. While they found a significant positive relationship between votes and Facebook likes, they concluded that their method was less accurate than using incumbency or traditional polling for figuring out whether a candidate would be elected.

Budiharto and Meiliana accurately predicted the Indonesian presidential election before the election occurred, tracking public opinion using Twitter data [13]. They collected tweets from presidential candidates and tweets containing the relevant hashtags for a combination of structural and sentiment analysis. While this paper did accurately predict the election, it did not publish performance evaluation metrics like MAE or $R^2$.

Awais et al. accurately predicted the 2018 General election in Pakistan using an array of different data sources from past election data and Twitter to approval polls [14]. Their model effectively predicted the winning candidates for every national assembly seat and predicted the seat share of political parties with 83% accuracy. For their Twitter data, they collected 640,000 tweets containing specific keywords and locations for three weeks before the election, utilizing a combination of sentiment and structural analysis to understand underlying support for the party. In sum, they concluded that "the right mixture of machine learning and artificial intelligence models could have a significant impact on the modern day political landscape" [14].

Sabuncu et al. is the first study attempting to predict the 2020 US presidential election with Twitter data [15]. It attempted to forecast election results using the number of positive, negative, and neutral tweets, ranking these tweets by their structural significance, which they determined to be the number of retweets. For their model, they used an autoregressive fractionally integrated moving average model (FARIMA) and 10 million tweets collected from September to November. They concluded that this method was more accurate than even polling data with a 1.52% MAE.

After a thorough review of the literature in the field, it is clear that there are a number of critical decisions which factor into the overall method for measuring public opinion from social media data. First, should one utilize data not from social media, and if so how should that information be used? Early research in this field focused only on using social media data but it appears that this decision was in part to establish that Twitter was a "forum for political deliberation" [3]. Recent research has found accurate methods for prediction involving polls and surveys to create "the right mixture" of data [14]. Since critical data for election prediction like aggregate polling metrics will not soon be replaced by these social media-based methods, it seems short-sighted to create a model that does not input this information. Second, what type of features should the model focus on from the social media data? From the accuracy of most recent studies and the conclusions of Skoric et al. [2], it seems clear that the combination of a structural and sentiment analysis provides the best accuracy as it utilizes more features from the data. Third, what type of analysis will be performed? This aspect of the model seems to have less impact than the data collected or the features analyzed, but still impacts the overall accuracy of the model. While supervised machine learning methods are common, this type of modeling is a time series problem and studies show that time series analysis provides accurate results -- sometimes more accurate than attribute-based models [11]. While these are the most important aspects of an electoral prediction model based on social media data, a fourth important aspect is the attention to key figures in the election process. Since social media data exhibits small-world properties, it is important to place emphasis on the most important figures and track the response to their use of social media [7].

This paper attempts to create a novel contribution to the field of measuring public opinion by performing a time series analysis on the sentiment and structural components of social media data while emphasizing the critical users and utilizing polling data. So that these results can be compared with others in the field, this paper will evaluate based on the MAE and $R^2$ standard established by Skoric et al. [2].

## 4. DATA

One of the biggest challenges of this project was to overcome and work around the scarcity of social media data. Though the preferred method was to perform prediction using multiple platforms, Twitter became the sole source for social media data as it was the easiest platform to collect data from and had the most publicly accessible datasets for this election. Even though Twitter has the most accessible data, it still provides its own challenges and obstacles.

Most free Twitter data using a developer account to access the Twitter API is only accessible as historical tweets from specific accounts or tweets collected in real-time while searching for keywords, hashtags, phrases, or account mentions. Twitter has recently opened up access to its full historical archive of tweets for academic researchers as of January 2021 but this access is not extended to undergraduate researchers. Unfortunately, this data roadblock was unknown during the course of the election (when I was still performing general research on the field and unsure of what data to collect) and thus limited the amount of data collected. For this project, four different social media datasets were evaluated for potential use but only two were used in the final analysis -- one publicly available online and one personally collected.

The first dataset utilized in this paper's analysis was a 1.7 million tweet data set from Kaggle user Manch Hui which scraped all tweets from 10/15-11/8 that included "#DonaldTrump", "#Trump", "#Biden", or "#JoeBiden" [16]. While this dataset only covered a limited amount of days before the election, it did publish all of the metadata for the tweets and collected tweets with common keywords. It is important to mention that this dataset is potentially biased as it only contains these four hashtags; there may be a large cohort of voters who would not tweet in favor of either candidate but rather use hashtags showing disdain for a certain candidate. Nevertheless, this dataset is this paper's source for tweets pertaining to the general conversations on Twitter for the 20 days before the election.

The second Twitter dataset used in this paper was from personally scraping historical tweets from multiple prominent accounts. This was done using the Twitter API and a python script which scraped the most recent tweets of 57 accounts. This dataset consists of the last 3200 tweets -- the maximum amount of recent tweets from a specific account -- of the presidential candidates and their running mates, 20 news organizations, 20 political pundits, and 20 popular politicians. The accounts of news organizations, political pundits, and politicians were all balanced so that there are 10 right-leaning accounts and 10 left-leaning accounts. These tweets were collected to potentially analyzeJürgens et al.'s [7] supposed small-world effects on Twitter. While only tweets from Trump and Biden's accounts were utilized in the final analysis, the goal for this dataset was to create two indexes of major liberal and conservative figures and measure the response to each over the course of the election.

One public dataset examined but not used was a 20 million tweet data set collected from 7/1-11/11 [15]. This dataset, found on IEEE Dataport, is composed of tweets that were searched by using party names, their abbreviations, candidates' names, and election slogans. While this dataset perhaps might have been useful, it did not contain all of the tweet metadata which prevented a thorough structural analysis. In addition, they collected tweets with a strange assortment of keywords and phrases that likely do not encompass the key political discussions happening on the platform.

The second data set found online but not used was a repository from a research article that collected 868 million tweets from 5/20/2019 to weeks after the election [17]. This was the only dataset found online which collected tweets that mentioned specific keywords and tweets from specific accounts. This dataset would have provided the perfect assortment of data for this project but unfortunately, it is nearly inaccessible. Although the tweet ids are posted on Github, these researchers were prohibited from publishing the entire tweet data due to Twitter's developer policy. Thus, the tweets must be rehydrated before they can be used. But rehydrating a tweet counts towards a monthly tweet cap usage, and a normal monthly cap is 500,000 tweets. To analyze and extract specific tweets from this dataset would require rehydrating hundreds of millions of tweets which was impossible with this project's resources.

In addition to social media data, this analysis incorporates the aggregate polling data of the two major candidates. This polling data is the forecasted variable in the time series model behind this analysis. To find accurate polling data, polling aggregators that provide robust estimates of public opinion were compared and RealClearPolitics aggregate was chosen. The data was acquired by scraping the RealClearPolitics average of polls from 9/31/2019 to the day of the election.

One very important aspect of this data and this project is its accessibility so that this thesis may be properly reviewed and reproduced. The main goal of this reproducibility plan is to provide as much information as possible for ease of analysis of these findings. All the code, outputs, and graphics have been uploaded to a Github repo so that the final version of this project can be analyzed and reproduced. That Github repo is linked [here](). As mentioned before, it is against Twitter's developer policy to post any data collected using the Twitter API except the id of tweets. Therefore, I will post the RealClearPolitics aggregate polling data but can only post the ids of tweets and not their metadata.

## 5. METHODOLOGY

As discussed previously, this paper attempts to combine previous successful methods in research to create a new method for analysis in measuring public opinion. The main goal for this analysis is to separately predict the vote share of Joe Biden and Donald Trump using a time series analysis converted to a supervised learning problem. There are two supervised learning models, one for Biden and one for Trump, each which process the Twitter data for a specific candidate and attempt to predict their vote share.

To begin, it is important to explain how the data was prepared before it was ingested by the model. This paper's method attempts to conduct a sentiment and structural analysis of the conversations on Twitter using Manch Hui's dataset of tweets that were scrapped because they contain a hashtag of the candidates [16]. These tweets were uploaded into data frames from two separate CSV files, one of which contained tweets with #Biden and #JoeBiden, and the other which contained tweets with #Trump and #DonaldTrump. The text of each tweet was analyzed using VADER (Valence Aware Dictionary and sEntiment Reasoner), a lexicon and rule-based sentiment analysis tool that is specifically designed for sentiments expressed on social media and trained on the text of tweets. While VADER does not give insight into the sentiment of media associated with the tweet and rather only analyses the text of the tweet, it does provide a good analysis of most of the tweets in the dataset. VADER provides a combined sentiment score that ranges from -1 to 1 with -1 as the maximum negative score and +1 as the maximum positive score. This gives insight into whether the tweet was talking about the candidate in a positive or negative light. One important caveat to the VADER score is that a simple range from -1 to 1 does not encapsulate many of the complex emotions mentioned in many political tweets. For example, a simple scale from negative to positive may obscure the overall sentiment of a tweet from a political supporter who simply is providing constructive criticism of policy. Then the combined VADER score was multiplied by the number of likes, retweets, and followers of the user who tweeted it. These structural features give context to how the message about the candidate was received by the platform and how many people may have even seen the tweet. In the last step, the sentiment-scored amount of likes, retweets, and followers were aggregated by the day summing the scores for each day. In total, 884,391 tweets were analyzed for sentiment and structural features from 20 days of Twitter activity.

This data regarding the general sentiment and breadth of conversations about candidates on Twitter was complemented by the tweets directly from candidates' accounts. These 1,181 tweets provided data on how the Twitter community was responding to the candidates. Over the same time period, the number of retweets and likes for each tweet were divided by the number of followers the candidate had. The tweets were aggregated by averaging the two ratios of likes to followers and retweets to followers by the day. Then this data was combined with the previous data from tweets with specific hashtags to generate two data frames (one for Biden and one for Trump) with an index of the days and these columns: Sum of Sentiment*Likes, Sum of Sentiment*Retweets, Sum of Sentiment*Followers, Mean of Biden Likes/Followers, and Mean of Biden Retweets/Followers.

Now that the data was properly formatted for the analysis, the next step was to convert a time series analysis to a supervised learning problem. This conversion was necessary as this prediction is mainly focused on the forecasting of polling data to predict a candidate's vote share. This polling data is a traditional time series dataset with metadata, the features from Twitter, describing each day. To include the metadata in the prediction of the next day's polling estimate, the model was converted to a supervised learning problem by using the sliding window method. This implementation of a sliding window simply added the previous day's polling estimate alongside each day's Twitter features so that the current day's polling estimate could be predicted from the current day's Twitter features and previous day's polling estimate. This first in this process involved checking to ensure each series or column was stationary. Stationary data is important for time series analysis as it enforces that the mean's expected value for the data is similar across different time periods. This helps ensure that the model does not vary in accuracy at any specific time point. To check each series for stationarity, an Augmented Dickey-Fuller unit root test from stats models was utilized with a significance level of 0.05. Removing stationarity helps make the mean and variance of the time series data consistent over time, thus making the time series easier to model. Examining each series concluded that six of the ten series were non-stationary and thus the entire dataset was differenced to make those series stationary. After, the same test was used to check each series again and revealed that eight of the ten series were then stationary. While optimally all series would be stationary, it would have taken seven differencings to reach this goal reducing the dataset from 19 rows with 80% stationarity to 13 rows with 100% stationarity. Due to the scarcity of data, the dataset was only differenced once.

The second step in converting this analysis to a supervised learning problem was to add a lagged variable using the sliding window method for the output of the previous time period. The previous day's aggregate polling metric was added to both of the dataframes. This was the final dataframe describing Biden that was used for the analysis:

| Date | Sum of Sentiment*Likes | Sum of Sentiment*Retweets | Sum of Sentiment*Followers | Mean of Biden Likes/Followers | Mean of Biden Retweets/Followers | Previous Polling Estimate |
|---|---|---|---|---|---|---|
| 2020-10-16 | 6.075432 | -0.883943 | -416.643181 | 0.000006 | 0.000069 | 51.7 |
| 2020-10-17 | -11.997177 | -1.605169 | 6.600909 | 0.000174 | -0.000176 | 51.2 |
| 2020-10-18 | -0.532162 | -0.191953 | 781.941740 | -0.000191 | -0.000577 | 51.3 |
| 2020-10-19 | 0.358707 | 0.119817 | 295.607700 | 0.001870 | 0.013707 | 51.3 |
| 2020-10-20 | 0.687242 | 0.297876 | -1223.989339 | -0.001564 | -0.010134 | 51.3 |

Figure 1.  Dataframe of Features Describing Biden's Vote Share

For prediction, the calculation of vote share for one day was utilized as the previous polling estimate for prediction of vote share in the next day. The training and test set splits were as close to an 80%/20% split as possible with the first 15 days being used for training the model and the last 4 days for evaluating its performance. Additionally, the last three days of the training set (again aiming for an 80%/20% split) were utilized as a validation set for choosing which algorithm to use. Multiple machine learning regression algorithms, discussed later, were trained on the first 12 days of the data. These algorithms inputted the features from the Twitter data and the previous day's polling estimate for each candidate separately and outputted the estimated polling of the current day. Multiple algorithms were trained on this training set of 12 days and were then evaluated on the next three days of data to see what algorithm had the lowest MAE. The algorithm with the lowest MAE would then be chosen for evaluation on the test set, the last four days of data. The performance on the test set is the performance reported and compared to appropriate baselines.

Because of the lack of data, significant algorithm exploration for which performed best would have overfit the data. More data would allow for algorithm selection to have a larger impact on refining the full model for the best possible accuracy. Five regression machine learning algorithms were used from scikit-learn's python package with the default hyperparameters; they were Lasso, Elastic-Net, Ridge Regression, and two Support Vector Regression (SVR) models, one with a linear kernel and one with an RBF kernel. All five algorithms were trained on the first 12 days of data and then evaluated on the next three to see which algorithm best fit the data. The linear SVR model performed best on the validation set, was trained on the entire training set, and subsequently was used for prediction on the test set.

## 6. FINDINGS

The findings for predicting the 2020 presidential election are based on three separate evaluation metrics. First in order of importance is whether the model predicted the correct winner. The second is the MAE of the predicted vote shares compared to the actual vote shares recorded for the election. The third is the average $R^2$ for each model's predictions when compared to aggregate polling numbers from the days in the test set. The best algorithm, an SVR model with a linear kernel, trained on the data from 10/16-10/30 performed well on two out of these three metrics. It correctly predicted the winner, Joe Biden, and recorded an MAE of 1.85%. The $R^2$ metric performed quite poorly likely due to the very small size of the test set and a best-fitting linear line would have fit the data better. Here are the predicted vote shares for each model compared to the aggregate polling numbers:

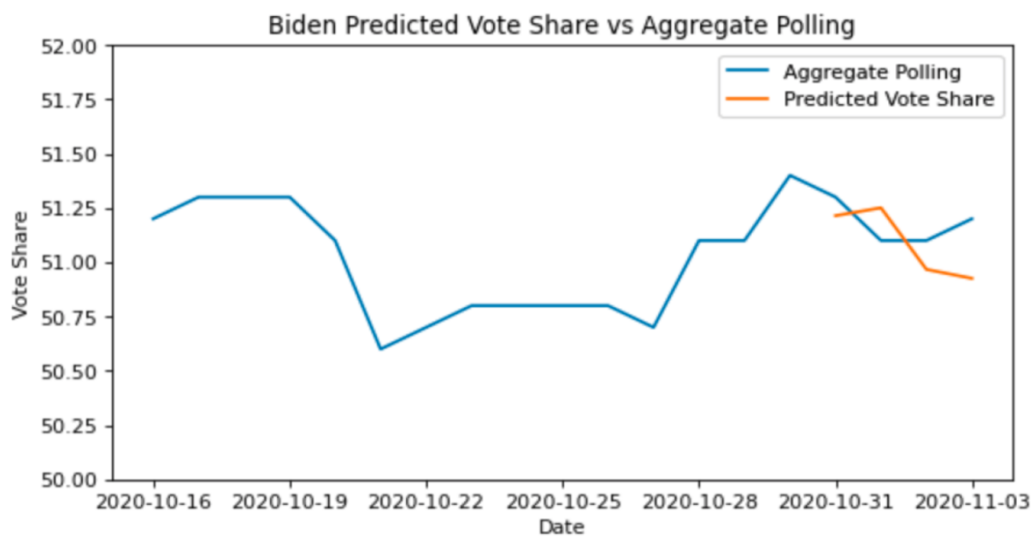

Figure 2. Graph of Prediction of Vote Share vs Aggregate Polling for Biden

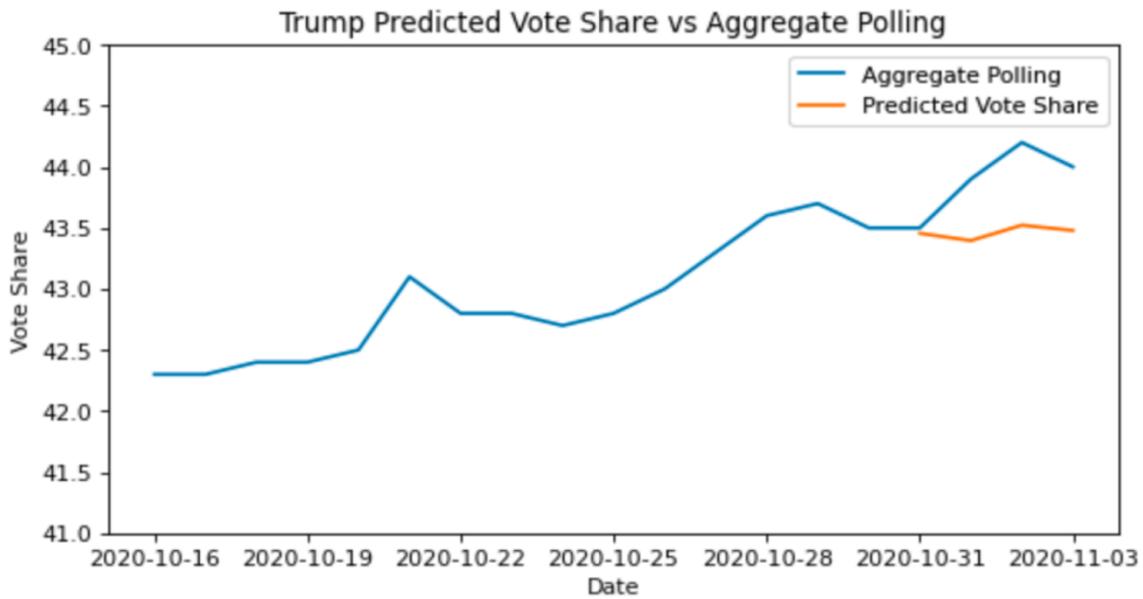

Figure 3. Graph of Prediction of Vote Share vs Aggregate Polling for Biden

Additionally, as the model was trained on polling data unlike most other studies in the field it is difficult to compare this paper's results with others. The influence of polling data gives this research an unfair advantage in comparison to research that did not utilize this type of data. Instead, a fair baseline for the performance of the model would be its comparison to the data it is attempting to perform better than, the aggregate polling data. While one cannot make a comparison based on $R^2$, it is possible to compare this model by the first two metrics. RCP's aggregate of polls does correctly predict the winner of the election and achieves an MAE of 1.5%, 0.35 less than this model. Thus while this method does achieve a lower MAE than the average MAE for lexi-sentiment (sentiment analysis with a lexicon-based approach) and structure models in research, it does not show an improvement in accuracy when evaluated to a comparable baseline [2].

## 7. CONCLUSION

With this data, this combination of approaches was not more successful in predicting the results of the 2020 presidential election than an appropriate baseline of the aggregate polling data. Despite that conclusion, I believe a more accurate summary of this research is that it is inconclusive as more data and more research is needed. With such a small sample size of data, only 20 days, it is very difficult to accurately determine if this method is truly effective. Thus, if this research were to be duplicated or expanded upon, my first recommendation would be to utilize more data. Optimistically, this project would have utilized data from early in the summer, around June, when most voters were aware of the likely Democratic and Republican nominees. Twitter data available from June onward would have provided months of data to train and evaluate a model with beyond sufficient test data. With this amount of data, one could conclusively determine whether these combinations of features from Twitter can perform better than the polling data.

Furthermore, there are four other important recommendations that could drastically improve the accuracy of this model. First, one should implement a better sentiment analysis method for determining the sentiment of a tweet. While the VADER analysis is sufficiently accurate, a machine learning model trained specifically on political tweets could provide much greater insight into each tweet's text. This change should greatly improve the model's ability to monitor conversations each day on Twitter. Second, other major political accounts could be monitored to provide even more data points for how the overall Republican and Democratic bases are reacting to conversations on Twitter. Following just the candidates' accounts biases the data to only reflect the support for those specific candidates, even though many voters vote based on party lines, not specific candidates. Tracking tweets from political pundits, other politicians, or news outlets like Fox News and MSNBC, could provide valuable data into the current enthusiasm of a party's base. While this was attempted in an earlier version of this research, accurately measuring the response to these major political accounts proved quite difficult to combine with the rest of the Twitter data.

A third recommendation for improving the accuracy of the model is to have multiple sources of social media data. As mentioned previously, Skoric et al. found that blogs scored better than Twitter for MAE and Facebook, forums, and blogs all reported better measures of $R^2$ [2]. Most notably, Skoric et al. reported that using multiple platforms for analysis reported substantially better measures of MAE. Thus, adding more social media data from multiple sources would lead to more conclusive positive results, showing if a technique is truly effective in predicting election results. Lastly, an additional way to validate whether the model was conclusively accurate would be to gather data from past presidential elections and predict their outcomes. This would help ensure that the model did not perform well on simply one election and rather that it is a robust method for predicting presidential elections. Comparison to old elections could also help validate which specific ideas, like sentiment analysis or a special focus on important users, are repeatedly effective.

If all of these changes were made, a model that performed better than the polling data in this one race could not conclusively be determined as an accurate model. It would also have to be used in forecasting future elections. As Huberty[18] shows in his research of multi-cycle election forecasting, performance in back-casting tests does not necessarily correlate with success in predicting future elections. This paper provides likely the best test for any model developed to forecast elections: can the forecasting method perform well for forward-looking predictions? In essence, can the method properly forecast in multiple election cycles? In order for a method to be recognized as successful, it should be tested on future election cycles and perform better than an adequate baseline, whether that is incumbency, polling predictions, or another metric.

In sum, this research proposes a new method for predicting elections that combines sentiment and structural data with aggregate polling, a time series analysis, and a special focus on Twitter users critical to the election. While this method performed worse than its baseline of polling predictions, it is inconclusive whether this is an accurate method for predicting elections due to the scarcity of data. More research and more data are needed to accurately measure this method's overall effectiveness.

**AUTHOR**

**Michael Caballero** recently graduated from the University of California, Berkeley with an Honors Data Science major and Computer Science minor.

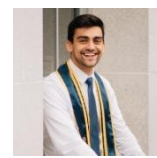